\PassOptionsToPackage{unicode}{hyperref}
\PassOptionsToPackage{hyphens}{url}
\PassOptionsToPackage{dvipsnames,svgnames,x11names}{xcolor}
\documentclass[
]{article}

\usepackage{amsmath,amssymb}
\usepackage{iftex}
\ifPDFTeX
  \usepackage[T1]{fontenc}
  \usepackage[utf8]{inputenc}
  \usepackage{textcomp} 
\else 
  \usepackage{unicode-math}
  \defaultfontfeatures{Scale=MatchLowercase}
  \defaultfontfeatures[\rmfamily]{Ligatures=TeX,Scale=1}
\fi
\usepackage{lmodern}
\ifPDFTeX\else  
\fi
\IfFileExists{upquote.sty}{\usepackage{upquote}}{}
\IfFileExists{microtype.sty}{
  \usepackage[]{microtype}
  \UseMicrotypeSet[protrusion]{basicmath} 
}{}
\makeatletter
\@ifundefined{KOMAClassName}{
  \IfFileExists{parskip.sty}{%
    \usepackage{parskip}
  }{
    \setlength{\parindent}{0pt}
    \setlength{\parskip}{6pt plus 2pt minus 1pt}}
}{
  \KOMAoptions{parskip=half}}
\makeatother
\usepackage{xcolor}
\makeatletter
\ifx\paragraph\undefined\else
  \let\oldparagraph\paragraph
  \renewcommand{\paragraph}{
    \@ifstar
      \xxxParagraphStar
      \xxxParagraphNoStar
  }
  \newcommand{\xxxParagraphStar}[1]{\oldparagraph*{#1}\mbox{}}
  \newcommand{\xxxParagraphNoStar}[1]{\oldparagraph{#1}\mbox{}}
\fi
\ifx\subparagraph\undefined\else
  \let\oldsubparagraph\subparagraph
  \renewcommand{\subparagraph}{
    \@ifstar
      \xxxSubParagraphStar
      \xxxSubParagraphNoStar
  }
  \newcommand{\xxxSubParagraphStar}[1]{\oldsubparagraph*{#1}\mbox{}}
  \newcommand{\xxxSubParagraphNoStar}[1]{\oldsubparagraph{#1}\mbox{}}
\fi
\makeatother

\usepackage{longtable,booktabs,array}
\usepackage{calc} 
\usepackage{etoolbox}
\makeatletter
\patchcmd\longtable{\par}{\if@noskipsec\mbox{}\fi\par}{}{}
\makeatother
\IfFileExists{footnotehyper.sty}{\usepackage{footnotehyper}}{\usepackage{footnote}}
\makesavenoteenv{longtable}
\usepackage{graphicx}
\makeatletter
\def\maxwidth{\ifdim\Gin@nat@width>\linewidth\linewidth\else\Gin@nat@width\fi}
\def\maxheight{\ifdim\Gin@nat@height>\textheight\textheight\else\Gin@nat@height\fi}
\makeatother
\setkeys{Gin}{width=\maxwidth,height=\maxheight,keepaspectratio}
\makeatletter
\def\fps@figure{htbp}
\makeatother
\NewDocumentCommand\citeproctext{}{}

\makeatletter
 \let\@cite@ofmt\@firstofone
 \def\@biblabel#1{}
 \def\@cite#1#2{{#1\if@tempswa , #2\fi}}
\makeatother
\newlength{\cslhangindent}
\newlength{\csllabelwidth}
\newenvironment{CSLReferences}[2] 
 {\begin{list}{}{%
  \setlength{\itemindent}{0pt}
  \setlength{\leftmargin}{0pt}
  \setlength{\parsep}{0pt}
  \ifodd #1
   \setlength{\leftmargin}{\cslhangindent}
   \setlength{\itemindent}{-1\cslhangindent}
  \fi
  \setlength{\itemsep}{#2\baselineskip}}}
 {\end{list}}
\usepackage{calc}

\usepackage{arxiv}
\usepackage{orcidlink}
\usepackage{amsmath}
\usepackage[T1]{fontenc}
\makeatletter
\@ifpackageloaded{caption}{}{\usepackage{caption}}
\AtBeginDocument{%
\ifdefined\contentsname
  \renewcommand*\contentsname{Table of contents}
\else
  \newcommand\contentsname{Table of contents}
\fi
\ifdefined\listfigurename
  \renewcommand*\listfigurename{List of Figures}
\else
  \newcommand\listfigurename{List of Figures}
\fi
\ifdefined\listtablename
  \renewcommand*\listtablename{List of Tables}
\else
  \newcommand\listtablename{List of Tables}
\fi
\ifdefined\figurename
  \renewcommand*\figurename{Figure}
\else
  \newcommand\figurename{Figure}
\fi
\ifdefined\tablename
  \renewcommand*\tablename{Table}
\else
  \newcommand\tablename{Table}
\fi
}
\@ifpackageloaded{float}{}{\usepackage{float}}
\floatstyle{ruled}
\@ifundefined{c@chapter}{\newfloat{codelisting}{h}{lop}}{\newfloat{codelisting}{h}{lop}[chapter]}
\floatname{codelisting}{Listing}

\makeatother
\makeatletter
\@ifpackageloaded{caption}{}{\usepackage{caption}}
\@ifpackageloaded{subcaption}{}{\usepackage{subcaption}}
\makeatother
\makeatletter
\@ifpackageloaded{algorithm}{}{\usepackage{algorithm}}
\makeatother
\makeatletter
\@ifpackageloaded{algpseudocode}{}{\usepackage{algpseudocode}}
\makeatother
\makeatletter
\@ifpackageloaded{caption}{}{\usepackage{caption}}
\makeatother

\ifLuaTeX
  \usepackage{selnolig}  
\fi
\usepackage{bookmark}

\IfFileExists{xurl.sty}{\usepackage{xurl}}{} 
\hypersetup{
  pdftitle={Enhancing robustness of Digital Shadow for CO2 Storage Monitoring with Augmented Rock Physics Modeling},
  pdfauthor={Abhinav Prakash Gahlot; Felix J. Herrmann},
  colorlinks=true,
  linkcolor={blue},
  filecolor={Maroon},
  citecolor={Blue},
  urlcolor={Blue},
  pdfcreator={LaTeX via pandoc}}

\newcommand{\runninghead}{A Preprint }
\title{Enhancing robustness of Digital Shadow for CO\textsubscript{2}
Storage Monitoring with Augmented Rock Physics Modeling}
\def\asep{\\\\\\ } 
\author{\textbf{Abhinav Prakash Gahlot}\\\\Georgia Institute of
Technology\\\\\asep\textbf{Felix J. Herrmann}\\\\Georgia Institute of
Technology\\\\}
\date{}
\begin{document}
\maketitle

\floatname{algorithm}{Algorithm}

\newcommand{\argmin}{\mathop{\mathrm{argmin}\,}\limits}
\newcommand{\argmax}{\mathop{\mathrm{argmax}\,}\limits}

\[
\def\textsc#1{\dosc#1\csod} 
\def\dosc#1#2\csod{{\rm #1{\small #2}}} 
\]

\section{Introduction}\label{introduction}

In order to achieve climate targets, the IPCC emphasizes the importance
of technologies capable of removing gigatonnes of CO\textsubscript{2}
each year, with Geological Carbon Storage (GCS) being a central
component of this strategy (Panel on Climate Change) 2018; Ringrose
2020). GCS involves capturing CO\textsubscript{2} and injecting it into
deep geological formations, where it can be safely stored over long
periods. To ensure the safety of GCS, accurate monitoring of subsurface
CO\textsubscript{2} behavior is essential, as it guarantees that
CO\textsubscript{2} plumes remain contained and do not leak into
surrounding formations (Ringrose 2023). Time-lapse seismic imaging has
proven vital for tracking the migration of CO\textsubscript{2} plumes,
yet it often fails to fully capture the complexities of multi-phase
subsurface flow.

Digital Shadows (DS), which utilize machine learning-driven data
assimilation techniques such as nonlinear Bayesian filtering and
generative AI (Spantini, Baptista, and Marzouk 2022; Gahlot, Orozco, et
al. 2024), offer a more detailed and reliable approach for monitoring
CO\textsubscript{2} storage (Herrmann 2023; Gahlot et al. 2023; Gahlot,
Li, et al. 2024; Gahlot, Orozco, et al. 2024). By incorporating
uncertainty in reservoir properties like permeability, this framework
improves the accuracy of CO\textsubscript{2} migration forecasts,
including plume pressure and saturation, and thereby reduces risks for
GCS projects. However, data assimilation depends on assumptions
regarding reservoir properties, rock physics models linking reservoir
state to seismic properties, and initial conditions. If these
assumptions are inaccurate, predictions may become unreliable, which in
turn can jeopardize the safety of GCS operations. One way to mitigate
this risk is by augmenting the forecast ensemble used to train the
neural networks responsible for the data assimilation process---mapping
prior forecast samples to the posterior. In this presentation, we
demonstrate that augmenting the forecast ensemble by incorporating
various rock physics models mitigates the negative effects of using
inaccurate models (e.g., uniform versus patchy saturation models).
Additionally, we find that ensemble augmentation can enhance predictive
accuracy in certain scenarios.

\section{Methodology}\label{methodology}

Building on the formulation in Gahlot, Orozco, et al. (2024) for
CO\textsubscript{2}-plume dynamics and observations, we introduce an
uncertainty-aware Digital Shadow framework, described by the following
equation:

\begin{equation}\phantomsection\label{eq-dynamics}{
\begin{aligned}
\mathbf{x}_k & = \mathcal{M}_k\bigl(\mathbf{x}_{k-1}, \boldsymbol{\kappa}_k; t_{k-1}, t_k\bigr)\\
              & = \mathcal{M}_k\bigl(\mathbf{x}_{k-1}, \boldsymbol{\kappa}_k\bigr), \ \boldsymbol{\kappa}_k \sim p(\boldsymbol{\kappa}) \quad \text{for}\quad k=1, \dots, K.
\end{aligned}
}\end{equation}

Here, \(\mathbf{x}_k\) represents the time-varying spatial distribution
of CO\textsubscript{2} saturation and pressure perturbations measured at
time steps \(k = 1, \dots, K\), while \(\mathcal{M}_k\) is the
multi-phase fluid-flow simulation operator that evolves the state from
time \(t_{k-1}\) to \(t_k\) for each time step. The symbol
\(\boldsymbol{\kappa}\) denotes the permeability distribution, which is
highly heterogeneous (Ringrose 2020, 2023). Since exact permeability
values are unknown, we assume a statistical distribution
\(p(\boldsymbol{\kappa})\) and sample from it during each timestep. The
state of CO\textsubscript{2} saturation and reservoir pressure
perturbations, \(\mathbf{x}_{k-1}\), is then propagated forward to
\(\mathbf{x}_k\).

While reservoir simulations provide accurate modeling of
CO\textsubscript{2} plume dynamics, relying solely on these simulations
is impractical due to the inherent stochasticity of permeability and the
uncertainty in rock physics models, which remain poorly constrained. To
improve the accuracy of CO\textsubscript{2} plume forecasts, monitoring
is key. Time-lapse seismic data (Lumley 2010) offer essential insights
into plume behavior, and we integrate these data into our framework to
enhance the characterization of the CO\textsubscript{2} plume. The
seismic data are modeled as:

\begin{equation}\phantomsection\label{eq-obs}{
\mathbf{y}_k = \mathcal{H}_k(\mathbf{x}_k;\mathcal{R}_k) + \boldsymbol{\epsilon}_k, \quad \boldsymbol{\epsilon}_k \sim p(\boldsymbol{\epsilon}), \quad \mathcal{R}_k \sim p(\mathcal{R}) \quad \text{for}\quad k=1, \dots, K
}\end{equation}

where \(\mathcal{H}_k\) is the observation operator, \(\mathbf{y}_k\) is
the seismic data observed at timestep \(k\), and
\(\boldsymbol{\epsilon}_k\) represents colored Gaussian noise added to
the seismic shot records before reverse-time migration. The symbol
\(\mathcal{R}_k\) denotes the rock physics model, randomly drawn from
the family of Brie Saturation models (Avseth, Mukerji, and Mavko 2010),
with the exponent \(e\) uniformly sampled from the range
\(e \sim \mathcal{U}(1,10)\). To combine fluid-flow simulations with
time-lapse seismic data, we use a nonlinear Bayesian filtering approach,
incorporating neural posterior density estimation to approximate the
posterior distribution of the CO\textsubscript{2} plume state
conditioned on seismic observations. This approach merges sequential
Bayesian inference with amortized neural posterior density estimation
(Papamakarios et al. 2021). To estimate the posterior distribution of
the CO\textsubscript{2} plume from seismic data, we employ Conditional
Normalizing Flows (CNFs) (Gahlot, Orozco, et al. 2024). In this
simulation-driven framework, we generate simulation pairs consisting of
CO\textsubscript{2} plume dynamics and corresponding seismic
observations (cf.~equations \ref{eq-dynamics} and \ref{eq-obs}) and
train the CNFs on these pairs. To account for different rock physics
models, we create multiple seismic images for each CO\textsubscript{2}
plume sample, resulting in an augmented dataset.

\section{Synthetic Case Study}\label{synthetic-case-study}

To evaluate our approach, we use a synthetic 2D Earth model derived from
the Compass model (E. Jones et al. 2012), representative of the North
Sea region. We focus on a subset of this model, containing subsurface
features suitable for CO\textsubscript{2} injection, which is
discretized into a grid of \(512 \times 256\) with a grid spacing of
\(6.25 \ \mathrm{m}\). To initialize the probabilistic digital shadow,
we require an ensemble of potential CO\textsubscript{2} plume scenarios.
Given the uncertainty in the permeability distribution of
CO\textsubscript{2} storage sites, we first establish a probabilistic
baseline for permeability, which we use to initialize the plume. The
permeability distribution is obtained through probabilistic
full-waveform inversion (Yin et al. 2024), assuming that a baseline
seismic survey was conducted prior to CO\textsubscript{2} injection.
This permeability distribution is then converted into permeability
samples using the empirical relationship from (Gahlot, Orozco, et al.
2024).

\subsection{Multi-Phase Flow
Simulations}\label{multi-phase-flow-simulations}

Flow simulations are carried out using the open-source tool JutulDarcy
\href{https://github.com/sintefmath/JutulDarcy.jl}{JutulDarcy.jl}
(Møyner, Bruer, and Yin 2023). Initially, the reservoir is filled with
brine, and supercritical CO\textsubscript{2} is injected at a rate of
\(0.0500 \ \mathrm{m^3/s}\) over a period of 1920 days. The injection
depth is approximately \(1200 \ \mathrm{m}\). The simulation is run over
four time-lapse intervals, \(t_k\), producing predicted
CO\textsubscript{2} saturation for each of the \(N=128\) ensemble
members at each timestep.

\subsection{Augmented Seismic
Simulations}\label{augmented-seismic-simulations}

The outputs from the 128 flow simulations are converted into changes in
subsurface acoustic properties using 10 different exponents of the Brie
Saturation model (Avseth, Mukerji, and Mavko 2010), thus augmenting the
data tenfold by generating different acoustic changes corresponding to
each flow simulation. Seismic surveys are conducted with 8 receivers and
200 sources, using a dominant frequency of 15 Hz and a recording
duration of 1.8 seconds. 28 dB SNR colored Gaussian noise is added to
the shot records. Nonlinear wave simulations and imaging are performed
using the open-source package
\href{https://github.com/slimgroup/JUDI.jl}{JUDI.jl} (Witte et al. 2019;
Louboutin et al. 2023).

\subsection{CNF Training}\label{cnf-training}

The \(1280\) ensemble members are used as training pairs, consisting of
forecasted CO\textsubscript{2} plumes and corresponding seismic
observations. We train a Conditional Normalizing Flow (CNF) using the
open-source package
\href{https://github.com/slimgroup/InvertibleNetworks.jl}{InvertibleNetworks.jl}
(Orozco et al. 2024). The network weights \(\boldsymbol{\phi}\) are
optimized by minimizing the following objective over 120 epochs using
the \(\textsc{ADAM}\) optimizer (Kingma and Ba 2014):

\begin{equation}\phantomsection\label{eq-loss-CNF}{
\widehat{\boldsymbol{\phi}} = \mathop{\mathrm{argmin}\,}\limits_{\boldsymbol{\phi}} \frac{1}{M}\sum_{m=1}^M \Biggl(\frac{\Big\|f_{\boldsymbol{\phi}}(\mathbf{x}^{(m)};\mathbf{y}^{(m)})\Big\|_2^2}{2} - \log\Bigl |\det\Bigl(\mathbf{J}^{(m)}_{f_{\boldsymbol{\phi}}}\Bigr)\Bigr |\Biggr).
}\end{equation}

where \(\mathbf{J}\) is the Jacobian of the network \(f_{\theta}\) with
respect to its input, and \(M\) is the number of training samples. For
further details, we refer to (Gahlot, Orozco, et al. 2024).

\section{Results}\label{results}

\begin{figure}

\begin{minipage}{\linewidth}

\centering{

\includegraphics{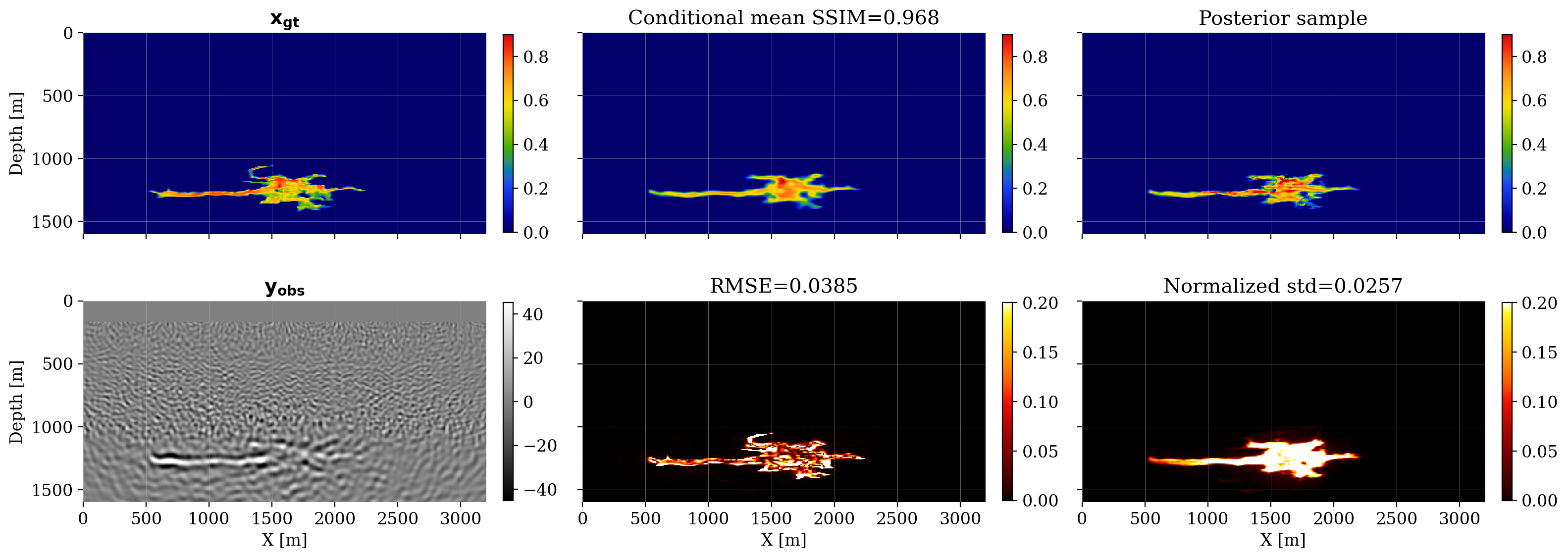}

}

\subcaption{\label{fig-naug-idy}Non-augmented, in-distribution y}

\end{minipage}%
\newline
\begin{minipage}{\linewidth}

\centering{

\includegraphics{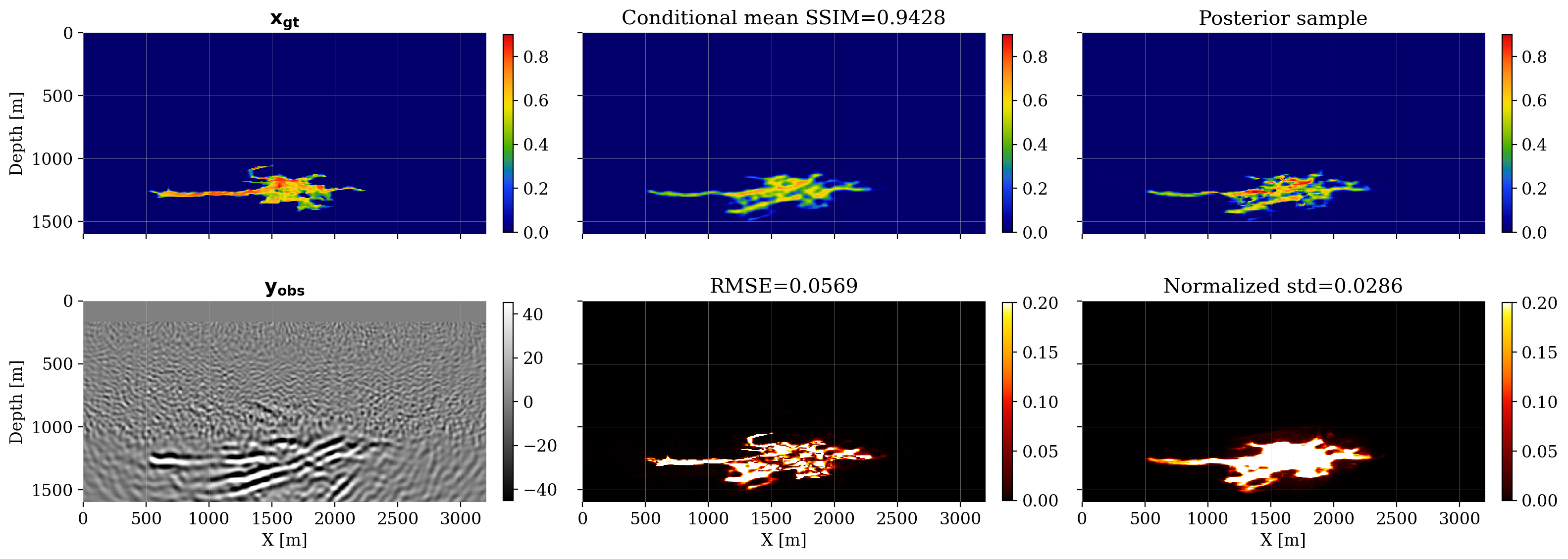}

}

\subcaption{\label{fig-naug-oody}Non-augmented, out-of-distribution y}

\end{minipage}%
\newline
\begin{minipage}{\linewidth}

\centering{

\includegraphics{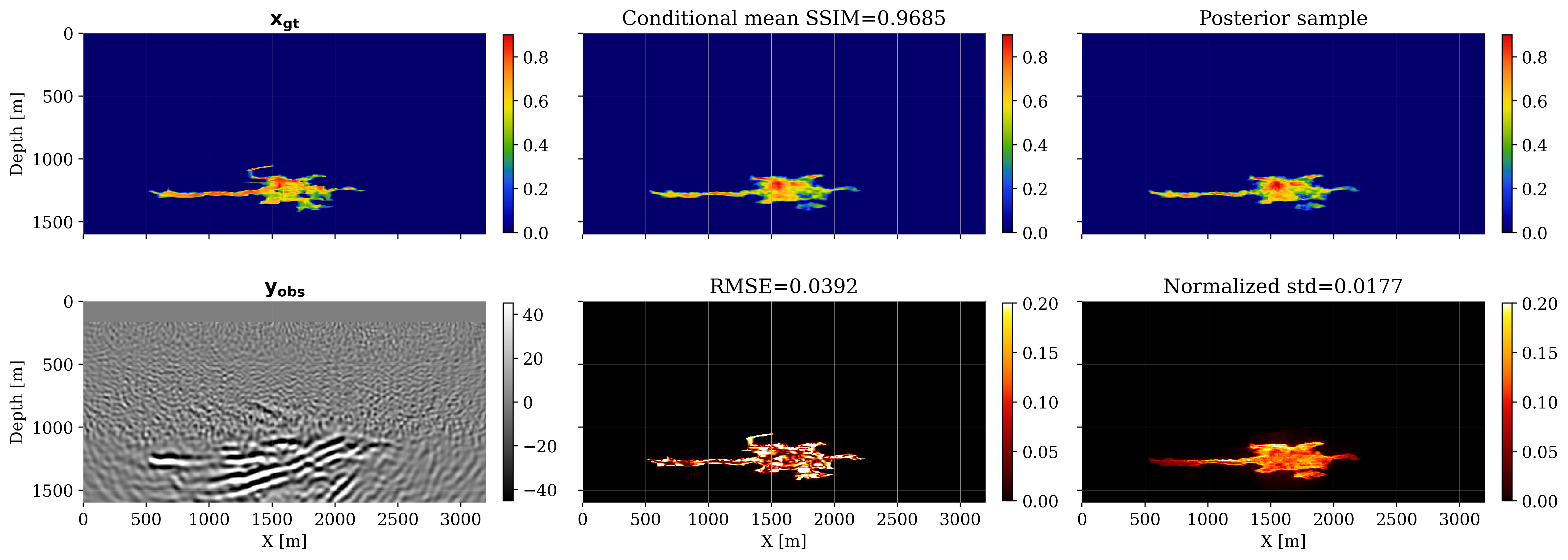}

}

\subcaption{\label{fig-aug-oody}Augmented, out-of-distribution y}

\end{minipage}%

\caption{\label{fig-sup}Inference results of digital shadow at \(k=1\)
\emph{(a)} without data augmentation and known rock physics model for
seismic observation \emph{(b)} without data augmentation and unknown
rock physics model for seismic observation \emph{(c)} with data
augmentation and unknown rock physics model for seismic observation.}

\end{figure}

The performance of our enhanced Digital Shadow framework, incorporating
rock physics, is shown in figure~\ref{fig-sup}. The top row in each
figure presents, from left to right, the ground truth (GT)
CO\textsubscript{2} plume, the conditional mean of the posterior
samples, and a sample from the posterior. The bottom row displays, from
left to right, the seismic observation, the error between the
conditional mean and the ground truth, and the uncertainty. As shown in
figure~\ref{fig-naug-idy}, the non-augmented DS performs well when the
seismic observation matches the correct rock physics model, with the
conditional mean closely approximating the GT. However, performance
degrades when the seismic observation is based on a different, unknown
rock physics model, as seen in figure~\ref{fig-naug-oody}, where the
conditional mean deviates from the GT, leading to increased error and
uncertainty. Augmenting the ensemble improves generalization when
seismic observations are based on an unknown rock physics model, as
demonstrated in figure~\ref{fig-aug-oody}. In this case, the conditional
mean is closer to the GT, and both error and uncertainty are reduced
compared to the non-augmented approach.

\section{Conclusions}\label{conclusions}

This study demonstrates that uncertainties in permeability and rock
physics models can significantly impact the accuracy of
CO\textsubscript{2} plume predictions. By augmenting the forecast
ensemble, DS is able to account for unknown rock physics models (at
least within the family of Brie saturation models where the exponent
\(e\) is not specified) and fluid-flow properties. The enriched dataset
improves the fidelity of CO\textsubscript{2} plume forecasts, thereby
enhancing the reliability of GCS monitoring.

\section{Acknowledgement}\label{acknowledgement}

This research was carried out with the support of Georgia Research
Alliance, partners of the ML4Seismic Center and in part by the US
National Science Foundation grant OAC 2203821. The overall readability
is enhanced using ChatGPT 4.

\clearpage
\section*{References}\label{references}
\addcontentsline{toc}{section}{References}

\phantomsection\label{refs}

\end{document}